\documentclass[12pt]{article}
\usepackage[T1]{fontenc}
\usepackage[utf8]{inputenc}  
\usepackage{graphicx}  
\usepackage{amsmath}   
\usepackage{xurl} 
\usepackage{lipsum}    
\usepackage{ragged2e}
\usepackage[authoryear]{natbib}
\usepackage[breaklinks]{hyperref}
\AtBeginDocument{\sloppy}

\title{\textbf{Decoding the Black Box: Integrating Moral Imagination with Technical AI Governance}}

\author{
  Krti Tallam \\
  EECS, University of California at Berkeley \\
  \texttt{ktallam@berkeley.edu}
}
\date{\today}

\begin{document}
\maketitle

\begin{abstract}
This paper examines the intricate interplay among AI safety, security, and governance by integrating technical systems engineering with principles of moral imagination and ethical philosophy. Drawing on foundational insights from \textit{Weapons of Math Destruction} and \textit{Thinking in Systems} alongside contemporary debates in AI ethics, we develop a comprehensive multi-dimensional framework designed to regulate AI technologies deployed in high-stakes domains such as defense, finance, healthcare, and education. Our approach combines rigorous technical analysis, quantitative risk assessment, and normative evaluation to expose systemic vulnerabilities inherent in opaque, black-box models. Detailed case studies—including analyses of Microsoft Tay (2016) and the UK A-Level Grading Algorithm (2020)—demonstrate how security lapses, bias amplification, and lack of accountability can precipitate cascading failures that undermine public trust. We conclude by outlining targeted strategies for enhancing AI resilience through adaptive regulatory mechanisms, robust security protocols, and interdisciplinary oversight, thereby advancing the state-of-the-art in ethical and technical AI governance.
\end{abstract}

\section{Introduction}
The rapid evolution of artificial intelligence (AI) has catalyzed both groundbreaking innovations and significant societal challenges. As AI technologies become increasingly integrated into critical aspects of modern life, their influence extends well beyond the confines of laboratories and tech companies, shaping the infrastructure of defense, finance, healthcare, and education. However, traditional approaches to AI safety and governance—characterized by simplistic, binary frameworks—are proving inadequate to address the nuanced and often unpredictable risks that these systems introduce.

Current regulatory and ethical frameworks tend to focus on technical metrics such as accuracy and efficiency, often overlooking the broader implications of AI decisions. These approaches neglect the complex interplay of social, cultural, and economic factors underlying AI deployments, ultimately resulting in policies that may fail to protect the public or even exacerbate existing inequalities \cite{oneil2016weapons}. In response to these shortcomings, this paper advocates for a holistic reimagining of AI regulation—one that is agile enough to respond to rapid technological changes and robust enough to manage multifaceted societal impacts.

Central to our proposal is the integration of interdisciplinary perspectives. We draw upon the foundational concepts of systems thinking \cite{meadows2008thinking} to understand AI as part of a larger socio-technical ecosystem, where feedback loops and leverage points can either mitigate or magnify risks. Moreover, we introduce the notion of \textbf{moral imagination}—the capacity to envision ethical possibilities that transcend conventional performance metrics—as a critical tool in rethinking governance \cite{jobin2019global}. By doing so, we emphasize that effective AI oversight must balance technical robustness with a deep sensitivity to ethical considerations and societal values.

In this paper, we explore how insights derived from key theoretical perspectives, as well as contemporary debates in AI ethics \cite{oneil2016weapons}, can be synthesized to form a multi-dimensional framework for regulating AI. This framework aims not only to safeguard against technical vulnerabilities and systemic failures but also to ensure that AI systems operate in ways that promote fairness, accountability, and the public good. Through this interdisciplinary approach, we contend that a governance model that is flexible, transparent, and ethically informed is essential for managing the complex challenges posed by AI in high-stakes domains.

\subsection{Motivation}
The integration of AI into everyday life has introduced unprecedented challenges, particularly regarding fairness, transparency, and accountability. As highlighted by \textit{Weapons of Math Destruction} \cite{oneil2016weapons}, opaque and unregulated algorithms can entrench and even amplify existing societal inequities. Meanwhile, systems thinking \cite{meadows2008thinking} offers a powerful perspective for analyzing the complex, interconnected feedback loops that drive these dynamics. Recent calls by thought leaders, such as Stephen Wolfram \cite{wolframAI}, emphasize the need for philosophical inquiry and interdisciplinary collaboration in reimagining AI governance \cite{jobin2019global}.

As AI security experts, we bring firsthand experience in identifying and mitigating vulnerabilities such as adversarial attacks, data poisoning, and model manipulation \cite{brundage2018malicious}. Our work has shown that technical failures in AI systems can quickly cascade into broader societal harms, reinforcing biases \cite{buolamwini2018gender} and undermining trust. This expertise fuels our commitment to developing a governance framework that not only strengthens the technical robustness of AI systems but also integrates ethical oversight and systemic resilience.

By combining our deep technical insights with interdisciplinary approaches \cite{holstein2019improving}, we advocate for a comprehensive regulatory model that addresses both the security challenges and the ethical dimensions of AI. This holistic perspective is crucial for safeguarding public trust and ensuring that AI technologies serve the collective good.

\subsection{Scope}
This paper investigates:
\begin{itemize}
    \item The limitations and inherent risks in current AI governance paradigms, with an emphasis on technical vulnerabilities and ethical oversights.
    \item Interdisciplinary approaches that integrate systems thinking, moral imagination, and ethical philosophy to address the multifaceted challenges of AI.
    \item Practical strategies and regulatory mechanisms for ensuring secure, transparent, and equitable AI deployment in critical sectors such as defense, finance, healthcare, and education.
\end{itemize}

\subsection{Structure of the Paper}
The paper is organized as follows:
\begin{itemize}
    \item \textbf{Section 2:} Examines key themes from algorithmic bias and AI systemic risks, with insights from \textit{Weapons of Math Destruction} and related studies.
    \item \textbf{Section 3:} Explores the role of systems thinking in AI governance, highlighting feedback loops, leverage points, and resilience mechanisms.
    \item \textbf{Section 4:} Analyzes philosophical and ethical considerations, including moral imagination, transparency, and AI’s societal implications.
    \item \textbf{Section 5:} Proposes a multi-dimensional regulatory framework integrating technical safeguards, ethical oversight, and systemic governance mechanisms.
    \item \textbf{Section 6:} Presents case studies on AI governance in high-stakes domains, including defense, finance, healthcare, and education.
    \item \textbf{Section 7:} Discusses AI security challenges, adversarial threats, and the role of interdisciplinary oversight in mitigating systemic vulnerabilities.
    \item \textbf{Section 8:} Concludes with future directions, emphasizing the need for adaptive governance frameworks and interdisciplinary collaboration.
\end{itemize}

\section{Key Themes from Weapons of Math Destruction}
Cathy O’Neil’s \textit{Weapons of Math Destruction} (WMD) critically examines the adverse societal impacts of opaque, unregulated algorithms. In this section, we distill and expand upon three central themes from her work, which serve as a foundational critique of current AI systems and their governance.

\subsection{Opacity and the Black-Box Problem}
One of the most persistent issues highlighted in WMD is the lack of transparency inherent in many AI systems. These algorithms often function as “black boxes” where the underlying logic, data inputs, and decision-making processes remain hidden from public view. This opacity is frequently a result of proprietary technologies and the sheer complexity of advanced machine learning models \cite{rudin2019stop}. Without clear insight into how these systems operate, it becomes exceedingly difficult for regulators, stakeholders, or affected individuals to scrutinize, challenge, or even understand the decisions being made. This lack of transparency is particularly problematic when decisions have significant real-world consequences, such as determining creditworthiness, assessing risk in defense applications, or influencing educational outcomes. The black-box nature of these systems hinders accountability, making it challenging to verify whether the outcomes are fair, unbiased, or legally compliant.

\subsection{Reinforcing Biases}
Another critical theme in WMD is the way in which AI systems can perpetuate and even amplify existing societal biases \cite{oneil2016weapons}. AI models are typically trained on historical data that reflect past human decisions and societal structures, which often include entrenched prejudices \cite{friedman1996bias, barocas2016big, noble2018algorithms}. When these biased datasets are used to train models, the algorithms can reproduce and reinforce these biases in their outputs. 

For instance, in defense or finance, historical data may lead to the over-representation of certain groups as high-risk, systematically disadvantaging them \cite{buolamwini2018gender}. This feedback not only entrenches inequity but also creates a self-perpetuating cycle: the biased outputs are fed back into the system during retraining, further cementing the bias in future iterations. Such reinforcement undermines the potential for AI to serve as an unbiased decision-maker and instead embeds historical inequities into automated systems, deepening systemic discrimination. Empirical investigations of algorithmic decision-making in domains like criminal justice and online search \cite{angwin2016machine, noble2018algorithms} further highlight how these feedback loops can entrench structural inequalities rather than mitigate them.

\subsection{Self-Fulfilling Feedback Loops}
The phenomenon of self-fulfilling feedback loops is perhaps one of the most insidious consequences of deploying opaque and biased AI systems. When an algorithm classifies a group as high-risk or unsuitable based on historical bias, subsequent systems often adopt and amplify that classification. This can result in a cascade of discriminatory outcomes where the initial bias becomes self-reinforcing. For instance, if an algorithm used in finance designates certain neighborhoods as risky based on past data, this may lead to reduced investment and fewer financial opportunities for residents, which in turn validates the algorithm’s original assessment \cite{barocas2016big, lum2016to}. Similarly, in educational or defense contexts, an early misclassification or biased prediction can set off a series of decisions that further disadvantage the affected group, reinforcing systemic disparities over time \cite{kleinberg2017inherent, angwin2016machine}. Breaking these self-reinforcing cycles requires not only technical interventions—such as regular audits and updates to the model—but also robust regulatory oversight that ensures accountability and transparency at every stage of the AI lifecycle.

\section{Applying Systems Thinking}
Systems thinking, as elaborated in Donella Meadows’ \textit{Thinking in Systems}, provides a holistic framework for analyzing AI ecosystems. Rather than viewing AI technologies as isolated tools, this approach encourages us to consider them as integral components of larger, interconnected socio-technical systems. By understanding the interplay of various elements and feedback loops within these systems, we can better predict how changes will propagate and design more effective interventions.

\subsection{Feedback Loops and System Behavior}
AI systems operate within complex networks characterized by feedback loops that influence their behavior over time:

\begin{itemize}
    \item \textbf{Reinforcing loops} are self-amplifying processes in which an initial change sets off a chain reaction that further intensifies the original effect. For example, if an algorithm classifies a particular demographic as high-risk, this designation may lead to decisions—such as increased scrutiny, reduced financial opportunities, or more frequent law enforcement interventions—that restrict opportunities for that group. These outcomes, in turn, generate new data that confirm and even exaggerate the initial classification. Over time, the model reinforces the bias in a self-perpetuating cycle, embedding these inequities into both the decision-making process and the training data of future iterations.

    \item \textbf{Balancing loops} act as countervailing mechanisms that work to restore stability by mitigating deviations from an equilibrium state. In the realm of AI governance, balancing loops might be implemented through regulatory feedback mechanisms or periodic audits designed to detect and correct biases. For instance, routine external reviews and continuous monitoring can adjust an AI model’s parameters in response to new data, ensuring that its outputs remain aligned with ethical standards. These interventions help moderate any unintended drift, ensuring that corrective actions are taken to counteract any potential reinforcing loops, thus maintaining a more balanced and equitable system.
\end{itemize}

Understanding these loops is crucial for developing regulatory frameworks that can preempt systemic failures. By identifying how small changes in data or decision criteria can trigger cascading effects, policymakers and engineers can design interventions that stabilize the system, mitigate unintended consequences, and promote fairness and accountability.

\subsection{Leverage Points}
Within any complex system, leverage points are strategic areas where targeted interventions can lead to significant improvements. For AI governance, identifying these leverage points allows for impactful changes with minimal disruption:
\begin{itemize}
    \item \textbf{Transparent Data Practices:} Instituting clear protocols for data collection, processing, and disclosure helps expose biases and errors early in the development cycle, enabling corrective actions before widespread deployment.
    \item \textbf{Diverse Stakeholder Engagement:} Involving a broad spectrum of stakeholders—including technologists, ethicists, regulators, and community representatives—ensures that AI systems are designed with multiple perspectives in mind. This collaboration fosters solutions that are sensitive to the ethical and societal implications of AI.
    \item \textbf{Routine Audits and Continuous Monitoring:} Regular, independent audits of AI systems can detect emerging biases, security vulnerabilities, and other systemic issues. Continuous monitoring facilitates timely interventions, ensuring that the AI remains aligned with ethical standards and societal values.
\end{itemize}
Focusing on these leverage points enables regulators and developers to induce positive systemic changes without resorting to radical, disruptive measures. By strategically intervening at these critical junctures, we can enhance the resilience of AI systems, ensuring that they function in ways that are transparent, equitable, and responsive to evolving societal needs.

\subsection{Systemic Resilience and Adaptation}
Resilience refers to the capacity of a system to withstand shocks and adapt to change. In the context of AI, resilience is not merely about preventing system failure in the face of external perturbations, but also about enabling AI governance frameworks to evolve as technology and societal expectations change. A resilient AI ecosystem is characterized by its ability to:

\begin{itemize}
    \item \textbf{Absorb Shocks:} AI systems must be designed with robust fail-safes that can absorb unexpected inputs, cyber-attacks, or other disturbances without cascading failures. This involves not only technical redundancy and backup protocols but also the capacity to isolate and contain issues as they arise.
    \item \textbf{Learn Continuously:} As new data and experiences accumulate, AI systems should be capable of self-assessment and iterative improvement. This necessitates mechanisms for regular audits, feedback loops, and updates that ensure models do not drift away from ethical or performance standards.
    \item \textbf{Adapt to New Conditions:} The rapid pace of technological change means that regulatory frameworks must be flexible and responsive. Dynamic policy mechanisms, such as adaptive licensing and real-time risk assessments, are critical for updating standards in line with emerging trends and vulnerabilities.
    \item \textbf{Maintain Equitability:} Resilience also involves safeguarding against systemic bias. As AI systems interact with evolving social dynamics, continuous monitoring is essential to ensure that the systems do not inadvertently reinforce or amplify existing inequities.
\end{itemize}

Resilient governance structures for AI must be dynamic, capable of self-correction, and responsive to both technical and societal changes. Such frameworks will help ensure that AI systems remain reliable and equitable over time, even as they scale and evolve in increasingly complex environments.

\section{Philosophical and Ethical Considerations for AI}
The ethical dimensions of AI extend far beyond quantitative performance metrics like accuracy or efficiency. At its core, AI technology poses fundamental philosophical questions about agency, responsibility, and the nature of decision-making. This section explores these broader considerations, which are integral to designing and deploying AI systems responsibly.

\subsection{Beyond Technical Metrics}
While conventional measures such as accuracy, precision, and speed are important, they do not fully capture the broader societal implications of AI decisions. AI systems do not operate in a vacuum—their outputs and decisions can have profound and lasting impacts on human lives, social structures, and ethical norms. For instance, an AI model used in risk assessment for defense might statistically optimize for threat detection, yet it could also inadvertently reinforce existing biases, resulting in disproportionate targeting of certain communities. Similarly, an AI system that allocates credit in finance may achieve high performance on conventional metrics while perpetuating historical inequalities by denying opportunities to marginalized groups.

To truly evaluate the efficacy and ethical standing of AI systems, we must extend our criteria beyond technical performance. New evaluation frameworks should incorporate dimensions that capture the real-world, human-centric effects of AI, ensuring that these systems promote fairness and social well-being. This involves assessing AI systems based on:

\begin{itemize}
    \item \textbf{Human Impact:} Measuring how AI decisions influence individual opportunities, livelihoods, and overall societal well-being. This includes understanding the long-term effects on communities and ensuring that AI applications contribute positively to human development.
    \item \textbf{Equity and Fairness:} Evaluating whether AI systems serve to mitigate or inadvertently exacerbate existing social and economic disparities. It is crucial to assess whether these systems distribute benefits equitably and do not reinforce historical biases or systemic discrimination.
    \item \textbf{Transparency and Accountability:} Ensuring that the processes and algorithms behind AI decisions are transparent and open to scrutiny. Clear documentation, explainable outputs, and defined lines of accountability are essential to build trust and allow for meaningful oversight, enabling stakeholders to hold developers and deployers responsible for adverse outcomes.
\end{itemize}

Incorporating these additional evaluation criteria will help ensure that AI systems are not only technically proficient but also ethically sound and socially beneficial.

\subsection{Moral Imagination in AI}
Central to the ethical deployment of AI is the concept of \textbf{moral imagination}—the capacity to envision alternative futures where technology not only optimizes processes but also contributes to human flourishing and social justice. Unlike purely technical considerations, moral imagination challenges us to look beyond immediate efficiency gains and to reflect on the broader societal consequences of AI. It calls for a rethinking of conventional problem definitions and inspires innovative solutions that align with deep-seated ethical values.

Moral imagination invites all stakeholders—developers, policymakers, ethicists, and end-users alike—to engage in a creative dialogue about the role of AI in society. This imaginative approach emphasizes that technological progress should not be measured solely by speed or accuracy but by its capacity to enhance quality of life, reduce inequality, and uphold human dignity. In practice, moral imagination encourages stakeholders to:

\begin{itemize}
    \item \textbf{Reframe Problems:} Rather than focusing exclusively on short-term efficiency or profit, stakeholders are urged to consider the long-term societal impacts of AI systems. This involves redefining problems to include considerations of community well-being, environmental sustainability, and social equity. For instance, an AI system designed for urban planning should not only optimize traffic flow but also promote accessible public spaces and reduce pollution.
    \item \textbf{Embrace Diversity of Perspectives:} Effective moral imagination requires integrating the values and experiences of all affected communities, particularly those who have historically been marginalized. By involving a wide range of voices—ranging from technical experts to community representatives—AI systems can be designed to address real-world needs more holistically and avoid reinforcing existing biases.
    \item \textbf{Challenge Established Norms:} Moral imagination encourages us to question prevailing assumptions and explore unconventional approaches. This might mean rethinking traditional data collection practices, critically evaluating existing algorithms, or proposing new metrics for success that prioritize ethical outcomes. Challenging the status quo can lead to transformative innovations that better serve society.
\end{itemize}

By cultivating moral imagination, developers and policymakers can create AI systems that are not only technologically advanced but also ethically robust and socially beneficial. This proactive, imaginative approach helps to ensure that AI contributes to a more just, equitable, and sustainable future, transforming technology into a genuine force for positive change.

\subsection{Philosophical Inquiry and Ethical Oversight}
Philosophical inquiry plays a pivotal role in addressing the deeper questions raised by AI technologies. It goes beyond the surface-level evaluation of algorithms by examining the fundamental nature of decision-making, ethics, and human values. This inquiry is essential for unpacking issues such as:

\begin{itemize}
    \item \textbf{Agency and Autonomy:} What does it mean for an AI system to act autonomously, and how should we balance human oversight with machine decision-making? Philosophical inquiry challenges us to delineate the boundaries between automated processes and human judgment, ensuring that machines enhance rather than supplant human agency.
    \item \textbf{Responsibility and Accountability:} When AI systems cause harm, determining who is accountable becomes a complex ethical dilemma. Is it the developers who created the algorithms, the operators who deploy them, or could the system itself bear responsibility? By exploring these questions, philosophical analysis helps to establish frameworks for accountability that protect individuals and society.
    \item \textbf{Justice and Fairness:} In an era of algorithmic decision-making, how do we define and measure fairness? Philosophical inquiry urges us to consider not only statistical parity but also the broader implications of justice in social contexts. It promotes the development of mechanisms that ensure AI contributes to social equity and does not perpetuate existing biases or create new forms of discrimination.
\end{itemize}

To translate these philosophical insights into actionable standards, ethical oversight mechanisms are crucial. Interdisciplinary ethics committees, external audits, and continuous evaluation processes serve as practical means to integrate these considerations into the lifecycle of AI systems. These oversight structures ensure that AI technologies are judged not only on technical performance but also on their alignment with broader moral and societal values.

By embedding philosophical inquiry and ethical oversight into AI development, we move beyond narrow technical measures. This integrated approach fosters the creation of AI frameworks that are both innovative and just, ultimately ensuring that technology serves the public good and upholds human dignity.

\subsection{The Need for Philosophical Inquiry}
Philosophical inquiry compels us to grapple with profound questions about the nature of intelligence, autonomy, and moral responsibility in the age of AI. It challenges us to ask not only how AI systems function, but also what values they embody and propagate. By integrating moral imagination into our evaluation of AI, we can envision alternative futures where human dignity, social justice, and ethical integrity take precedence over mere computational efficiency. In this context, philosophers and ethicists play a critical role: they help delineate which outcomes are desirable and which practices may be morally unacceptable, guiding the development of AI technologies that reflect our highest ethical aspirations \cite{wolfram2024philosophers,taddeo2018how}.

\subsection{Beyond Technical Metrics}
Traditional evaluations of AI systems have largely concentrated on quantifiable performance indicators such as accuracy, precision, and speed. While these metrics are essential for assessing technical performance, they fall short of capturing the full impact of AI on human lives. The deployment of AI in areas that affect everyday living—be it in defense, finance, healthcare, or education—requires that we also account for considerations of social welfare, justice, and human dignity. A comprehensive evaluation framework must integrate these ethical dimensions, ensuring that AI technologies are designed and implemented in ways that uphold the broader interests of society and protect the rights of all individuals.

\subsection{Reimagining Human-AI Collaboration}
Conventional narratives often cast the relationship between humans and AI in adversarial or competitive terms, suggesting a dichotomy between man and machine. Instead, we propose a paradigm shift towards a collaborative model, where AI is viewed as a tool that augments human decision-making rather than replacing it \cite{shneiderman2020human, green2019principles}. In this envisioned framework, AI systems operate under strict ethical guidelines and are subject to continuous human oversight. This approach not only ensures that AI contributions are beneficial and aligned with societal values, but also fosters an environment in which human creativity and critical judgment work in tandem with technological innovation. By reimagining human-AI collaboration as a partnership, we can harness the strengths of both to address complex societal challenges in a more inclusive and equitable manner.

\section{Proposed Multi-Dimensional Framework for AI Regulation}
Building on the insights from the previous sections, we propose a comprehensive framework for AI regulation that integrates technical, ethical, and systemic perspectives. This framework is designed to ensure that AI systems are developed, deployed, and maintained in ways that are transparent, adaptive, ethically grounded, and capable of withstanding systemic challenges. The framework is organized into four key pillars:

\subsection{Transparency and Explainability}
Transparency forms the bedrock of ethical AI, as it allows stakeholders to understand how decisions are made and to hold systems accountable. Our framework emphasizes:
\begin{itemize}
    \item \textbf{Mandatory Model Documentation:} Developers must provide clear, detailed disclosures of the data sources, underlying assumptions, and limitations of their models. This documentation should include the rationale behind design choices and any known biases in the training data, enabling independent verification and facilitating public scrutiny.
    \item \textbf{Third-Party Auditing:} Regular audits by independent experts are essential to verify that AI systems adhere to both ethical guidelines and technical standards. These audits serve as an external check, ensuring that models remain compliant over time and that any deviations or emerging issues are promptly identified and addressed.
\end{itemize}
By prioritizing transparency and explainability, our framework fosters trust among users and regulatory bodies, paving the way for informed decision-making and accountability.

\subsection{Dynamic Regulatory Processes}
Given the rapid evolution of AI technologies, static regulatory frameworks quickly become outdated. Our proposal calls for a dynamic, agile regulatory approach that evolves in tandem with technological advancements:
\begin{itemize}
    \item \textbf{Feedback-Driven Legislation:} Regulations should be designed to incorporate real-time data and stakeholder feedback. This allows policies to be updated continuously as new challenges emerge, ensuring that governance remains relevant and effective.
    \item \textbf{Continuous Risk Assessments:} Regular evaluations must be conducted to identify potential vulnerabilities or adverse impacts. By implementing a system of ongoing risk assessments, regulators can adjust measures proactively, mitigating risks before they escalate into systemic failures.
\end{itemize}
Dynamic regulation ensures that as AI systems grow in complexity and reach, governance structures remain robust, responsive, and capable of safeguarding public interests.

\subsection{Ethical Review Boards and Philosophical Oversight}
Embedding ethical considerations into every stage of AI development is critical. Our framework proposes the creation of structured oversight mechanisms that integrate diverse perspectives:
\begin{itemize}
    \item \textbf{Interdisciplinary Panels:} Establish ethics committees composed of ethicists, philosophers, technologists, legal experts, and community representatives. These panels are tasked with evaluating AI systems against a broad set of ethical criteria, ensuring that the design and deployment of AI technologies align with societal values.
    \item \textbf{Ethics-by-Design:} Ethical principles should be integrated into the development lifecycle from the outset. This means that every stage—from conceptualization and design to deployment and maintenance—must include checkpoints to assess potential ethical dilemmas and societal impacts.
\end{itemize}
This proactive ethical oversight not only helps prevent harm but also drives innovation by challenging developers to explore alternative solutions that promote fairness and inclusivity.

\subsection{Systems-Level Monitoring and Intervention}
A systemic approach to AI governance involves continuous monitoring and timely intervention at the level of entire AI ecosystems:
\begin{itemize}
    \item \textbf{Identifying Leverage Points:} Regular assessments should be conducted to pinpoint critical junctures within AI systems where minor interventions could yield significant improvements. These leverage points are essential for preemptively addressing vulnerabilities before they propagate through the system.
    \item \textbf{Adaptive Licensing:} Implement regulatory mechanisms where AI systems are subject to periodic re-approval based on comprehensive evaluations of their performance, ethical compliance, and societal impact. This adaptive licensing model ensures that systems are continually re-assessed and updated in accordance with evolving standards and new insights.
\end{itemize}
Systems-level monitoring provides a macro-level safeguard, ensuring that AI systems function reliably and ethically within their broader socio-technical context.

In summary, this multi-dimensional framework for AI regulation is designed to address the complex challenges posed by modern AI systems. By combining transparency, dynamic regulation, ethical oversight, and systems-level monitoring, the framework aims to create an environment where AI technologies can thrive in a manner that is both innovative and aligned with the public good.

\section{Case Studies and Domain Applications}
The following case studies illustrate how our proposed framework can be applied across various domains.

\subsection{Defense}
AI systems in defense range from autonomous weapons to threat detection algorithms and strategic analysis tools. These high-stakes applications introduce significant ethical and security concerns that require a nuanced regulatory approach:

\begin{itemize}
    \item \textbf{Escalation Risks:} Autonomous systems may misinterpret adversary behavior or malfunction under stress, potentially triggering unintended military escalations \cite{scharre2018army}. Without robust oversight, such miscalculations could lead to rapid and uncontrollable conflict dynamics.
    \item \textbf{Bias in Targeting:} AI-driven threat analysis tools often rely on historical data, which can embed cultural or regional biases. This may result in skewed threat assessments, where certain populations or regions are unfairly targeted based on flawed or incomplete data \cite{freedman2019future}.
    \item \textbf{International Oversight:} The global nature of defense and security necessitates collaborative governance structures that transcend national borders. Establishing shared ethical standards and cooperative oversight mechanisms is critical for preventing unilateral actions that could destabilize international security \cite{un2021report}.
\end{itemize}

An illustrative example of addressing these challenges is provided by the Cybersecurity and Infrastructure Security Agency (CISA) under the Department of Homeland Security (DHS). CISA's role in safeguarding the nation's critical infrastructure offers valuable lessons for regulating AI in defense:
\begin{itemize}
    \item \textbf{Risk Management and Incident Response:} CISA has developed comprehensive protocols to anticipate, mitigate, and respond to cybersecurity threats. These protocols are essential for ensuring that AI systems in defense can identify and neutralize vulnerabilities in real time, reducing the risk of catastrophic failures.
    \item \textbf{Collaborative Governance:} CISA actively promotes cross-sector collaboration among government agencies, private industry, and international partners. This approach fosters a collective defense strategy that ensures AI systems are developed and managed under unified ethical and security standards.
    \item \textbf{Integration of Ethical and Security Standards:} By advocating for transparency, rigorous testing, and clear accountability, CISA’s guidelines help bridge the gap between technical performance and ethical oversight. This dual focus is crucial for ensuring that AI systems not only defend against cyber threats but also operate within frameworks that minimize bias and promote fairness.
\end{itemize}

In summary, the defense sector exemplifies the necessity for a multi-dimensional regulatory framework that integrates technical safeguards with ethical oversight and international collaboration. Drawing on the example of DHS CISA, it is evident that robust cybersecurity practices combined with interdisciplinary governance can help manage the inherent risks of AI in defense, ensuring that these systems contribute to security without compromising ethical standards or escalating conflicts.

\subsection{Finances}
Financial services represent one of the most significant arenas for AI application, spanning automated credit scoring and loan approvals to high-frequency algorithmic trading. However, these systems often rely on historical data that can embed longstanding biases. For instance, historical lending practices, which may have been influenced by discriminatory policies, can lead AI models to systematically deny credit to certain demographic groups. This not only reinforces existing economic disparities but also limits access to opportunities for underserved communities.

Moreover, algorithmic trading systems operate in highly dynamic environments where rapid, real-time decisions are the norm. In such settings, even minor anomalies or unanticipated shifts in market conditions can trigger cascading failures, potentially leading to significant market volatility and systemic risks \cite{kirilenko2017flash}. The rapid pace and complexity of these systems demand robust oversight mechanisms to ensure stability and fairness.

To mitigate these challenges, financial AI systems must embrace a regulatory framework that emphasizes transparency, accountability, and inclusivity. Key components of such a framework include:

\begin{itemize}
    \item \textbf{Rigorous, Independent Audits:} Regulatory bodies should require that credit scoring and loan approval models undergo thorough audits by independent experts. These audits must assess not only technical performance but also the ethical implications of model design and data usage, ensuring that biases are identified and rectified \cite{fuster2020predictably}.
    \item \textbf{Multi-Factor Evaluation:} Beyond conventional credit histories, AI models should integrate diverse data sources—such as alternative financial behaviors, employment records, and socio-economic indicators—to offer a more comprehensive and equitable assessment of creditworthiness.
    \item \textbf{Real-Time Monitoring Systems:} For algorithmic trading, continuous real-time monitoring is essential. Advanced anomaly detection systems can identify irregular patterns early, allowing for timely intervention to prevent cascading failures and mitigate systemic financial risks.
\end{itemize}

By adopting these measures, the financial sector can harness the benefits of AI to improve efficiency and decision-making while also promoting social equity and economic stability. This approach ensures that AI applications in finance contribute to an inclusive economic landscape, reducing the risk of adverse large-scale impacts while enhancing overall market integrity.

\subsection{Healthcare}
AI applications in healthcare—ranging from diagnostic imaging and patient prioritization to predictive analytics for disease management—offer transformative potential for improving patient outcomes. For instance, deep learning models have achieved high accuracy in interpreting radiological images, with systems like Google's DeepMind demonstrating promise in diagnosing eye diseases such as diabetic retinopathy \cite{gulshan2016development}. Similarly, AI-driven predictive models can forecast disease progression and tailor treatment plans, potentially revolutionizing personalized medicine.

However, these innovations also bring significant risks if not accompanied by rigorous oversight and regulation. A primary concern is the presence of biases in medical data. Historical datasets often reflect long-standing disparities in healthcare access and quality, leading AI systems to produce unequal treatment outcomes. A well-known example is the study by Obermeyer et al. (2019), which revealed that a widely used healthcare algorithm systematically underpredicted the health needs of Black patients compared to white patients, thereby limiting access to additional care for minority populations \cite{obermeyer2019dissecting}. Such biases in data can inadvertently reinforce and amplify existing inequities.

In addition to data bias, healthcare AI systems are vulnerable to adversarial attacks. Even minor, deliberate modifications to diagnostic images or patient records can mislead AI algorithms, resulting in dangerous misdiagnoses or inappropriate treatment plans. For example, research by Finlayson et al. (2019) demonstrated that subtle adversarial perturbations in medical imaging could cause a model to overlook critical features, such as tumors, leading to delayed treatments and poorer outcomes \cite{finlayson2019adversarial}. Moreover, security vulnerabilities in these systems can expose sensitive patient data, increasing the risk of breaches that compromise patient privacy and erode trust.

Together, these challenges highlight the need for a robust regulatory framework that combines ethical oversight with stringent technical safeguards. Such a framework would ensure that AI in healthcare is developed and deployed in ways that maximize benefits while minimizing risks, ultimately promoting equitable and secure patient care.

To address these challenges, healthcare AI must be governed through a robust framework that prioritizes both ethical oversight and strong security measures. Key strategies include:

\begin{itemize}
    \item \textbf{Dedicated Bias Monitoring Committees:} In practice, this could involve establishing a formal committee within a healthcare institution that meets regularly to evaluate AI systems. For example, a large hospital might form a committee composed of clinicians, data scientists, and ethicists tasked with reviewing the performance of a risk-scoring algorithm used for patient triage. By analyzing error rates and outcomes across different demographic groups—such as comparing how the algorithm performs for different racial or socioeconomic populations—the committee can identify biases that may be disadvantaging certain groups. This approach is similar to the methodology employed in studies like Obermeyer et al. (2019), which uncovered significant bias in healthcare risk assessments. Corrective actions might include recalibrating the algorithm or adjusting data inputs to ensure fairer treatment across all demographics.

    \item \textbf{Stringent Ethics Review Processes:} A real-world implementation might require that every new healthcare AI tool undergoes a thorough ethical review by an interdisciplinary ethics board before it is deployed. For instance, prior to rolling out an AI-driven diagnostic tool, a hospital's ethics board—which could include oncologists, AI experts, legal advisors, and patient representatives—would review the system's intended use, its potential risks, and its alignment with clinical best practices. This process could prevent issues like those experienced by IBM Watson for Oncology, where insufficient clinical validation and ethical oversight led to unsafe treatment recommendations. The ethics review process ensures that all potential impacts, including unintended consequences, are considered and mitigated early on.

    \item \textbf{Continuous Audits and Security-by-Design:} Continuous auditing involves setting up systems for ongoing monitoring and periodic review of AI systems to detect performance degradation, bias, or security vulnerabilities. For example, a healthcare network might mandate quarterly security audits of its AI tools used in radiology, incorporating adversarial testing methods that simulate potential attacks on the system. This could help uncover subtle vulnerabilities that might allow for adversarial manipulation, which, if unaddressed, could lead to dangerous misdiagnoses. Embedding security-by-design means that from the earliest stages of development, the AI system is engineered with robust security features—such as input validation, encryption of sensitive data, and regular penetration testing—to preempt potential exploits. This proactive approach ensures that both technical and ethical standards are maintained continuously throughout the system’s lifecycle.
\end{itemize}

By adopting these proactive measures, healthcare providers can harness the potential of AI to enhance diagnostic accuracy and treatment efficiency while ensuring that these technologies serve all patient populations equitably. A well-regulated, ethically grounded, and secure healthcare AI ecosystem is essential for realizing the full promise of AI in medicine and maintaining public trust in these transformative technologies.

\subsection{Education}
AI-driven applications in education—ranging from automated admissions and grading systems to personalized learning platforms—offer the potential to enhance educational efficiency and individualize instruction. Yet, these technologies also risk perpetuating historical inequalities if their design and deployment are not carefully managed. For example, algorithms that rely on historical performance data may inadvertently disadvantage students from under-resourced schools, thereby reinforcing cycles of inequity \cite{williamson2021education}.

The 2020 UK A-Level grading fiasco serves as a cautionary tale, demonstrating how an algorithmic system, when deployed without adequate oversight, can produce biased outcomes that undermine public trust \cite{walker2020algorithm}. To prevent such occurrences, educational AI systems must be designed with transparency at their core. This includes the implementation of clear, accountable scoring mechanisms and the regular reassessment of algorithmic outputs. Moreover, engaging diverse stakeholder perspectives—ranging from educators and students to community representatives—ensures that the AI systems reflect a broad spectrum of needs and values. By instituting rigorous external audits and establishing continuous feedback loops, the education sector can harness the benefits of AI while promoting fairness and providing equitable opportunities for all learners.

\subsection{Case Studies: Microsoft Tay and UK A-Level Grading Algorithm}
To further illustrate the practical implications of the challenges discussed in this paper, we now examine two high-profile case studies. These cases are particularly relevant as they highlight different facets of AI vulnerability: one demonstrates how insufficient security and oversight can transform an innovative AI tool into a propagator of harmful content, while the other reveals the social and ethical ramifications of deploying opaque, automated decision-making systems in high-stakes environments. Together, they underscore the urgent need for robust, multi-dimensional frameworks for AI regulation that integrate technical, ethical, and systemic perspectives.

\subsubsection{Microsoft Tay (2016)}
Microsoft Tay was an AI-powered Twitter chatbot designed to learn from interactions and emulate conversational behaviors. Launched in March 2016, Tay was intended to engage with users in a friendly, playful manner. However, within hours, malicious actors exploited vulnerabilities in its learning algorithm. By inundating Tay with offensive language and provocative content, users were able to manipulate its responses, leading the chatbot to produce and amplify hate speech. The rapid degeneration of Tay's output serves as a stark example of how a lack of robust security measures—such as safeguards against data poisoning and adversarial inputs—can derail an AI system. Tay's failure not only tarnished Microsoft's reputation but also highlighted the broader risks associated with deploying AI in uncontrolled, real-world environments without proper oversight and ethical guardrails \cite{hern2016microsoft}.

\subsubsection{UK A-Level Grading Algorithm (2020)}
In response to the COVID-19 pandemic, the UK government deployed an algorithm to predict A-level exam results after traditional examinations were canceled. Designed to standardize grades, the algorithm relied heavily on historical data and statistical models to determine student performance. However, the system disproportionately disadvantaged students from under-resourced schools, whose historical performance data did not reflect their individual potential or recent improvements. The resulting grades sparked widespread public outrage and protests, as many students and educators argued that the algorithm perpetuated systemic inequalities and failed to account for contextual factors affecting student performance. This case underscores the dangers of opaque, automated decision-making in high-stakes settings such as education. It emphasizes the necessity for adaptive, transparent, and accountable AI governance frameworks that can accommodate the complex realities of human contexts and ensure equitable outcomes \cite{walker2020algorithm}.

\section{AI Security, Safety, and Governance: A Systemic Perspective}

AI security is not merely a technical challenge but is deeply intertwined with ethical governance and systemic resilience. As AI systems become integral to critical decision-making processes, vulnerabilities in these systems can precipitate far-reaching consequences that extend beyond isolated technical failures \cite{brundage2018malicious}. A breach in security can undermine public trust, exacerbate existing biases, and ultimately destabilize the social fabric, making robust oversight a societal imperative \cite{bostrom2014superintelligence}.

\subsection{Security Vulnerabilities Fueling Bias and Feedback Loops}
Modern AI systems face a spectrum of security threats that can cascade into broader ethical and societal harms:
\begin{itemize}
    \item \textbf{Adversarial attacks} exploit subtle weaknesses in AI models. For instance, by introducing carefully crafted perturbations into input data, adversaries can cause a diagnostic algorithm to misinterpret medical images, potentially leading to dangerous misclassifications. In high-stakes domains like autonomous driving or defense, such attacks can result in catastrophic failures, illustrating how minor vulnerabilities can be weaponized with severe consequences.
    \item \textbf{Data poisoning} involves the deliberate injection of biased or malicious data into the training set. This can skew the model's outputs by embedding systematic errors from the outset. A real-world example includes instances where biased historical data in credit scoring models leads to the persistent underestimation of creditworthiness for marginalized communities. Over time, these biased outputs reinforce pre-existing inequities, making it difficult to break the cycle of discrimination.
    \item \textbf{Model manipulation} refers to the intentional alteration of a model's behavior through external interventions. Such manipulation can initiate self-reinforcing feedback loops where the model's biased outputs are continuously used to retrain the system, further entrenching discriminatory practices. For example, if a judicial risk assessment tool is manipulated to unfairly classify certain groups as high-risk, the resulting decisions may lead to harsher sentencing and further justify the biased algorithm, thus creating a vicious cycle of injustice.
\end{itemize}

These vulnerabilities underscore that failures in AI security are not isolated technical issues; they can trigger cascading effects that amplify systemic biases and ethical lapses. Consider a scenario in which an adversarial attack on an AI-driven autonomous vehicle causes it to misinterpret road signals—this could lead not only to accidents but also to a broader erosion of trust in automated transportation systems \cite{eykholt2018physical}. Similarly, data poisoning in financial algorithms might result in widespread denial of credit to minority groups, exacerbating economic disparities and undermining the integrity of financial markets \cite{jagielski2018manipulating}.

In light of these challenges, a systemic approach to AI governance is essential—one that integrates robust technical safeguards with continuous ethical oversight and adaptive regulatory measures. By addressing both the technical vulnerabilities and the broader ethical implications of AI, we can work towards creating resilient systems that uphold fairness, accountability, and public trust.

\subsection{Impact Across Critical Domains}
The consequences of AI security failures are profound and ripple across multiple critical sectors:

\begin{itemize}
    \item \textbf{Defense:} In the defense sector, insecure AI can lead to misdirected autonomous actions or the compromise of sensitive intelligence data. For example, if an adversary manipulates an AI system responsible for threat assessment or autonomous weapon control, the result could be erroneous target identification or unintended military actions. Such vulnerabilities risk escalating conflicts or triggering unforeseen retaliatory measures, thereby destabilizing national and international security.
    
    \item \textbf{Finance:} In finance, the impact of insecure AI manifests in several ways. Credit scoring and loan approval systems that rely on flawed or manipulated data can lead to biased decisions, denying credit to individuals from historically marginalized communities. Similarly, algorithmic trading systems are highly sensitive to even minor perturbations; a small malfunction or a coordinated adversarial attack could set off a chain reaction, leading to market instability or even a financial crisis. These failures not only harm individual livelihoods but can also undermine the overall integrity of financial markets.
    
    \item \textbf{Healthcare:} Security breaches in healthcare AI can have life-or-death consequences. For instance, if a diagnostic tool is compromised, it might produce erroneous results, leading to misdiagnoses or inappropriate treatment recommendations. Such failures are particularly dangerous for vulnerable communities who may already face disparities in access to quality healthcare. In addition, breaches that expose sensitive patient data can erode trust in medical institutions, hindering the broader adoption of potentially life-saving technologies.
    
    \item \textbf{Education:} In the education sector, insecure AI systems, such as those used for automated admissions or grading, can reproduce and even amplify historical biases. For example, if an algorithm that determines student placements is fed biased historical data, it may unjustly favor or disadvantage certain groups. This can lead to systemic discrimination, limiting academic opportunities for students from under-resourced backgrounds and perpetuating long-term inequities in educational outcomes.
\end{itemize}

\subsection{Towards Resilient AI Ecosystems: Interdisciplinary Security Frameworks}
Addressing AI security challenges requires an interdisciplinary approach that combines technical, ethical, and systemic perspectives. Such an approach ensures that vulnerabilities are not only identified but also remediated in a way that aligns with broader societal values.

\begin{itemize}
    \item \textbf{Philosophical and Ethical Foundations:} The creation of robust, secure AI systems must be guided by strong ethical principles. This involves embedding values such as fairness, justice, and accountability into the design and deployment processes. By engaging philosophers and ethicists in the development lifecycle, organizations can better anticipate the societal impacts of AI and ensure that systems are built to serve the public good.
    
    \item \textbf{Systems Thinking:} Applying systems thinking allows us to understand the interdependencies within AI ecosystems. It helps identify critical leverage points where small interventions can have a large impact on overall system behavior. For example, by mapping the data flows and decision pathways in a financial AI system, regulators can pinpoint where biases are likely to be introduced and design targeted interventions to mitigate these effects.
    
    \item \textbf{Multi-Stakeholder Governance and Public Accountability:} Effective AI security requires a governance model that brings together diverse stakeholders—government agencies, industry leaders, academic researchers, and community representatives. Such collaborative oversight ensures continuous monitoring, transparent reporting, and prompt remediation of vulnerabilities. Mechanisms like regular public audits, standardized reporting protocols, and adaptive regulatory frameworks help maintain accountability and foster public trust in AI systems.
\end{itemize}

By integrating these interdisciplinary security frameworks, we can build resilient AI ecosystems capable of withstanding both technical attacks and systemic ethical challenges. This holistic approach not only protects critical domains from immediate risks but also ensures that AI technologies evolve in a manner that is secure, equitable, and aligned with societal values.

\section{Future Directions}
\subsection{Summary of Insights}
Throughout this paper, we have argued that the challenges posed by AI are multifaceted, encompassing technical vulnerabilities, ethical dilemmas, and systemic risks. Our analysis underscores several key insights: opaque and biased algorithms, as highlighted in critiques such as \textit{Weapons of Math Destruction}, can exacerbate social inequities; systems thinking reveals the intricate feedback loops that may lead to systemic failures; and the integration of moral imagination and philosophical inquiry provides a pathway to reframe AI governance. This holistic perspective is essential to ensure that AI systems are both secure and equitable, serving the broader public interest rather than reinforcing existing disparities.

\subsection{Looking Ahead}
Future research and development in AI governance must address several critical areas to keep pace with rapid technological evolution:
\begin{itemize}
    \item \textbf{Adaptive Regulatory Processes:} There is an urgent need for regulatory frameworks that are flexible and responsive to technological advancements. Future work should explore mechanisms for real-time policy updates, such as dynamic licensing, continuous risk assessment, and adaptive standards that evolve alongside emerging AI capabilities.
    \item \textbf{Enhanced Interdisciplinary Collaboration:} Solving the complex challenges of AI governance requires contributions from diverse fields. Collaborative efforts among technologists, ethicists, policymakers, legal experts, and representatives from affected communities will be crucial. Establishing cross-sector partnerships and fostering platforms for dialogue can facilitate the development of standards and best practices that reflect a comprehensive understanding of both technical and societal dimensions.
    \item \textbf{Globally Harmonized Standards:} AI technologies operate across national boundaries, making international cooperation essential. Future research should focus on creating globally harmonized standards for AI safety, security, and governance. This includes developing common ethical frameworks, shared regulatory benchmarks, and collaborative oversight mechanisms that can help mitigate risks and ensure consistent application of best practices worldwide.
\end{itemize}

\subsection{Call to Action}
To create an AI ecosystem that truly benefits society, it is imperative that we fundamentally reframe our approach to governance. This means moving beyond outdated, binary regulatory models and embracing a holistic perspective that integrates technical robustness, ethical sensitivity, and systemic resilience. We call upon researchers, policymakers, industry leaders, and civil society to:
\begin{itemize}
    \item Champion the development and adoption of adaptive, forward-thinking regulatory frameworks that can keep pace with the rapid evolution of AI technologies.
    \item Foster interdisciplinary collaborations that bridge technical innovation with ethical and societal considerations, ensuring that diverse perspectives are represented in the governance process.
    \item Advocate for the establishment of internationally coordinated standards and oversight mechanisms, recognizing that the challenges posed by AI are global in nature and require collective action.
\end{itemize}

By embracing these proactive measures, we can transform AI into a catalyst for empowerment rather than a tool of harm, ensuring its benefits are shared equitably by all. The future of AI governance depends on our resolve to integrate robust safeguards and steadfast ethical principles into every stage of technology development, ultimately forging a resilient and just digital ecosystem for everyone.

\clearpage

\bibliographystyle{plainnat}
\bibliography{references}

\end{document}